\documentclass[showpacs,twocolumn,showkeys,amsmath,amssymb,pra,subscriptaddress,longbibliography,nofootinbib]{revtex4-1}    
\usepackage{amsmath,amssymb,amsfonts}   
\usepackage{graphicx}
\usepackage{txfonts}
\usepackage{epstopdf}
\usepackage[T1]{fontenc}
\usepackage[unicode]{hyperref}
\hypersetup{
   unicode=true,          
   a4paper=true,
   plainpages=false,
   pdftitle={Title of PDF},    
   pdfauthor={Author of PDF},     
   pdfsubject={Subject of PDF},   
   colorlinks=true,       
}
\begin{document}

\title{
Temperatures are not useful to characterise bright-soliton
  experiments for ultra-cold atoms}

\author{Christoph Weiss} 
\email{christoph.weiss@durham.ac.uk}


\affiliation{%
Joint Quantum Centre (JQC) Durham--Newcastle, Department
  of Physics, Durham University, Durham DH1 3LE, United Kingdom}

\author{Simon A. Gardiner}
\email{s.a.gardiner@durham.ac.uk}
\affiliation{%
Joint Quantum Centre (JQC) Durham--Newcastle, Department of Physics, Durham University, Durham DH1 3LE, United Kingdom}

\author{Bettina Gertjerenken}
\email{b.gertjerenken@uni-oldenburg.de}
\affiliation{%
Department of Mathematics and Statistics, University of Massachusetts, Amherst, MA 01003-4515, USA} 


\date {28 October 2016}

 \begin{abstract}
Contrary to many other translationally invariant  one-dimensional models, the low-temperature
phase for an attractively interacting
one-dimensional Bose-gas (a quantum bright soliton) is stable
against thermal fluctuations. However, treating the thermal
properties of quantum bright solitons within the canonical ensemble
leads to anomalous fluctuations of the total energy that indicate that
canonical and micro-canonical ensembles are not equivalent. State-of-the-art experiments are
best described by the micro-canonical ensemble, within which we predict a co-existence between single atoms
and solitons even in the thermodynamic limit --- contrary to strong predictions
based on both the Landau hypothesis and the canonical ensemble.
This questions the use of
 temperatures to
describe state-of-the-art bright soliton experiments that currently load
Bose-Einstein condensates into quasi-one-dimensional wave guides
without adding contact to a heat bath.

\end{abstract}


\maketitle

The experimental realization~\cite{MeyrathEtAl2005,GauntEtAl2013} of a box potential opens new doors in
investigating translationally invariant  systems of ultra-cold
atoms. For ideal gases in one-dimension there is no Bose-Einstein
condensate (BEC), whereas  the presence of a harmonic trap leads to a
quasi-condensate~\cite{HerzogOlshanii1997}. For attractively interacting
Bose gases, the ground state is a weakly bound molecule, a
matter-wave bright soliton. Some of its properties are remarkably different
from those of BECs: as we will see below, there is no off-diagonal long-range
order. Thus, mathematical
theorems about the non-existence~\cite{MerminWagner1966,Hohenberg1967}
of off-diagonal long-range order do not
lead to additional physical insight for this model system.

Matter-wave bright solitons have been
experimentally generated for ultracold
atomic
gases in quasi-one-dimensional wave guides for attractive
interactions~\cite{KhaykovichEtAl2002,StreckerEtAl2002,CornishEtAl2006,MarchantEtAl2013,MedleyEtAl2014,McDonaldEtAl2014,NguyenEtAl2014,EverittEtAl2015,MarchantEtAl2016,LepoutreEtAl2016}
and, in the presence of an optical lattice,  also for repulsive
interactions~\cite{EiermannEtAl2004}.  For
dark solitons~\cite{BurgerEtAl1999}, developing a complete understanding by
modeling them on the many-particle quantum
level~\cite{GirardeauWright2000,MishmashCarr2009,DelandeSacha2014,KronkeSchmelcher2014,KarpiukEtAl2015,SyrwidSacha2015}
is crucial. The same is true for bright
solitons. We note that \textit{quasi}-one-dimensional wave guides provide
thermalisation mechanisms~\cite{MazetsSchmiedmayer2010}. 

When modeling the statistical physics, experiments with ultra-cold atoms arguably are best
described by the micro-canonical ensemble (MCE, with fixed total energy $E$ and fixed
particle number $N$~\cite{PathriaBeale2011}). As long as the different ensembles are
equivalent one can choose a simpler ensemble, such as the canonical
ensemble (CE, which allows the energy to fluctuate; while the particle
number $N$ still is fixed, we now have the temperature $T$ as a
thermodynamic variable~\cite{PathriaBeale2011}). For attractive bosons
in a quasi-one-dimensional wave guide the canonical ensemble even
predicts that the thermal weakly-bound
molecule--non-molecule crossover becomes
a phase transition in the thermodynamic limit~\cite{HerzogEtAl2014,Weiss2016}. However, as we will
see, the energy
fluctuations are anomalously large, making it necessary to
re-investigate the transition on the level of the micro-canonical
ensemble. Furthermore, this implies that
while the canonical ensemble is a powerful tool to describe both the
high- and low-temperature phases~\cite{HerzogEtAl2014,Weiss2016}, it fails to capture the physics of
the crossover itself correctly for thermally isolated systems with
ultracold atoms.

For $N$ attractively interacting atoms ($g_{\rm 1D}<0$) in one
dimension, the integrable Lieb-Liniger-(McGuire)
Hamiltonian~\cite{LiebLiniger1963,McGuire1964} is a very useful model
\begin{equation}\label{eq:LL}
\hat{H} = -\sum_{j=1}^N\frac{\hbar^2}{2m}\frac{\partial^2}{\partial
  x_j^2}+\sum_{j=1}^{N-1}\sum_{n=j+1}^{N}g_{\rm 1D}\delta(x_j-x_n),
\;\; 
-\frac L2 \le x_j\le \frac L2,
\end{equation}
where $x_j$ denotes the position of particle $j$ of mass $m$, and the interaction strength
\begin{equation}
\label{eq:g1d}
g_{\rm 1D} =2\hbar
\omega_{\perp}a <0 
\end{equation}
 is set by the
\textit{s}-wave scattering
length $a$ and the perpendicular angular trapping-frequency
$\omega_{\perp}$~\cite{Olshanii1998}. For sufficiently large
$L$~\cite{footnoteWeissGardinerGertjerenken2016}
 we have the ground state energy~\cite{McGuire1964} 
\begin{equation}
\label{eq:E0}
E_0(N) =-\frac1{24}\frac{mg_{1\rm D}^2}{\hbar^2}N(N^2-1)\sim -\frac1{24}\frac{mg_{1\rm D}^2N^2}{\hbar^2}N,
\end{equation}
and all excited energies~\cite{CastinHerzog2001}
\begin{equation}
  \label{eq:AllEigenvalues}
 E_{\rm total}= \sum_{r=1}^R\left(E_0(N_r)+N_r \frac{\hbar^2k_r^2}{2m}\right),\;\; \sum_{r=1}^R N_r=N,\;\; N_r \ge 1,
\end{equation}
corresponding to the intuitive picture of solitonlets of size $N_r$
and their respective center-of-mass kinetic energies.

The ground state wave function
\begin{equation}
\label{eq:Psi0}
\Psi(x_1,x_2,\ldots,x_N) \propto \exp\left(-\frac{m |g_{\rm 1D}|}{2\hbar^2}\sum_{j<\nu}|x_j-x_{\nu}|\right);
\end{equation}
is translationally invariant and corresponds to a quantum bright
soliton --- a weakly bound molecule with delocalized center-of-mass wave
function and localized relative wave function. It also helps to
quantify what a sufficiently large $L$ is~\cite{footnoteWeissGardinerGertjerenken2016}.
The size of the
molecule can also be characterized by the single particle density (after
replacing the delocalized center-of-mass wave function by a delta
function at $X_0$~\cite{CalogeroDegasperis1975,CastinHerzog2001},
which is identical to
the mean-field result~\cite{PethickSmith2008})
\begin{equation} 
\label{eq:SolitonDensityLength}
|\varphi(x)|^2 \propto \frac 1{\cosh[(x-X_0)/(2\xi_N)]^2}, \quad
\xi_N\equiv \frac{\hbar^2}{m(N-1)\left|g_{\rm 1D}\right|}.
\end{equation}

In order to ensure that a finite-temperature phase transition does not
violate the (Hohenberg)-Mermin-Wagner
theorem~\cite{MerminWagner1966,Hohenberg1967}, 
 we use the thermodynamic limit~\cite{HerzogEtAl2014,Weiss2016} 
\begin{equation}
\label{eq:limit}
N\to\infty,\;\; L\to\infty,\;\; g_{\rm 1D}\to 0, 
\; \frac NL = \textrm{const.},\; \xi_N = \textrm{const.},
\end{equation}
where the particle number $N$, system length $L$,
interaction strength $g_{\rm 1D}$ and soliton length $\xi_N$ are
defined in Eqs.~(\ref{eq:LL}), (\ref{eq:g1d}) and (\ref{eq:SolitonDensityLength}).

 We note that there is no off-diagonal long-range
order\footnote{There is,
  however,  long-range order in the limit investigated in
  addition to the limit~(\ref{eq:limit}) in Ref.~\cite{HerzogEtAl2014}. Thus, given the (Hohenberg-)Mermin-Wagner theorem~\cite{MerminWagner1966,Hohenberg1967}
  one should expect the characteristic temperature to approach zero
  for the infinite system. This is indeed the case~\cite{Castin2014}.}
 for bright solitons in the
limit~(\ref{eq:limit}). The many-particle ground state can be viewed as consisting of a
relative\footnote{The wave-function described in terms of the relative
motional degrees of freedom --- which we call the \textit{relative
  wave-function.}} wave-function given by a Hartree product state with $N$
particles occupying the mean-field-bright-soliton mode $\varphi(x)\propto
\cosh[(x-X_0)/(2\xi_N)]^{-1}$ [see Eq.~(\ref{eq:SolitonDensityLength})]; and a center-of-mass wave function for the
variable $X_0$
(cf.~\cite{CalogeroDegasperis1975,CastinHerzog2001}). The one-body
density matrix~\cite{PitaevskiiStringari2003} then is $\propto
\cosh[(x-X_0)/(2\xi_N)]^{-1}\cosh[(x'-X_0)/(2\xi_N)]^{-1}$ which vanishes
in the limit $|x-x'|\to\infty$ even after integrating over
$X_0$. Thus, there is no off-diagonal long range order in our system~\cite{Weiss2016};
the existence of bright solitons at low but non-zero temperatures in
the thermodynamic limit~(\ref{eq:limit}) therefore does not violate the
(Hohenberg-)Mermin-Wagner
theorem~\cite{MerminWagner1966,Hohenberg1967}. 

However, there is a severe problem with the canonical description of
the attractive Lieb-Liniger gas in 1D: the scaling of the specific
heat~\cite{Weiss2016}. 
The average energy $\langle E\rangle$ changes sign at the transition
temperature from a negative value~(\ref{eq:E0}) --- which scales
$\propto N$ in the limit~(\ref{eq:limit}) --- to a
positive value $\propto N$
within a temperature range $\delta T\propto 1/N$. By using the
textbook results for calculating the specific heat~\cite{PathriaBeale2011} we
thus obtain
\begin{equation}
\label{eq:CNsq}
C_{L,N}\propto \Delta E^2 \propto \frac{\partial \langle E\rangle}{\partial T}\approx
\frac{\delta E}{\delta T}\propto N^2
\end{equation}
near the transition temperature. 
Ensemble equivalence between MCE and CE would usually require a
scaling not faster than $N$ (rather than $N^2$) --- as is the case
outside the transition region~\cite{Weiss2016}.  Thus a complete description will
require a \textit{micro-canonical treatment} with fixed total energy
(possibly including some very small energy uncertainty)
within the MCE, which is the most suitable ensemble for
state-of-the-art bright soliton
experiments~\cite{KhaykovichEtAl2002,StreckerEtAl2002,CornishEtAl2006,MarchantEtAl2013,MedleyEtAl2014,McDonaldEtAl2014,NguyenEtAl2014,EverittEtAl2015,MarchantEtAl2016,LepoutreEtAl2016}. Nevertheless
we can profit from the CE result as we know what happens both at high
temperatures ($N$ single atoms) and at low temperatures (a single
bright soliton containing all $N$ atoms)~\cite{HerzogEtAl2014,Weiss2016}. We only have to identify
the energy regimes which correspond to these two phases. Note that the
disagreement between MCE and CE only happens in the transition region
of size $\delta T$ which vanishes in the limit~(\ref{eq:limit}) and thus appears to be negligible for practical
purposes. However, as we will see below, this region covers a huge
energy scale and covers all experimentally accessible cases of
realising bright solitons in a one-dimensional wave guide --- as long
as 3D effects can be discarded.

For very low total energies [$E_{\rm total} \gtrsim
 E_0(N)$], when
approaching the internal ground state energy [$E_{\rm total}\to E_0(N)$], only the internal ground state
(plus some non-degenerate kinetic energy of the big soliton) is
energetically accessible --- as long as
\begin{equation*}
E_0(N-1) > E_{\rm total}.
\end{equation*} 
For slightly higher (less negative) total energy, the only
energetically accessible states are the $N$-particle soliton or,
alternatively, an $N-1$-particle soliton and a single unbound particle. Naively
applying the Landau hypothesis\footnote{\label{foot:LandauHypthesisText}According to the textbook series by Landau and Lifshitz (the very end
of Ref.~\cite{LandauLifshitz2002b}), the
co-existence of single unbound atoms with one bound state or several
bound states is
excluded in a large class of 1D systems. } suggests that it must be the former as
the only ``allowed'' options seem to be either the $N$ single atoms or
an $N$-particle soliton. However, when counting possible
configurations the $N$-particle soliton will always lose compared to
distributing the kinetic energy among two or more smaller solitons or
solitons and single particles.

We hasten to
add that there is of course no error in the derivation of the Landau
hypothesis (footnote~\ref{foot:LandauHypthesisText}; page \pageref{foot:LandauHypthesisText}) as given in what
is frequently considered to be an authoritative text on  
theoretical physics~\cite{LandauLifshitz2002b}. However, its powerful statement is based on
assumptions that, while fulfilled by a huge class of models, are not
fulfilled by the
attractive 
Lieb-Liniger model. One example is that the size of many objects
scales with the number of particles $N$ the object is composed of; like
neutron stars, bright solitons
become smaller with larger particle numbers
[Eq.~(\ref{eq:SolitonDensityLength})].

Here, as long as the total energy is negative, $0>E_{\rm total}$,
within the MCE the configuration consisting of $N$ single atoms is simply
not energetically accessible (the
$N$ single atom case has a positive total energy) and the configuration of an $N$-atom
soliton is in most cases statistically irrelevant. A summary of the
three energy regimes can be found in Table~\ref{table:overview}.
\begin{center}
\begin{table}[t]
    \begin{tabular}{| c || c| c | c |} 
\hline
    ~~regime~~ & ~$E_{\rm total}\gtrapprox 0$~  &
 ~$0\gtrapprox E_{\rm  total}>E_0(N-1)$~ & ~$E_0(N-1)> E_{\rm total}$~ \\\hline
        MCE & $N$ atoms &coexistence & one large soliton\\ \hline
     \end{tabular}
\caption{For positive total energies, the micro-canonical ensemble
  (MCE) agrees with the canonical prediction that there are $N$ single
atoms~\cite{HerzogEtAl2014,Weiss2016}. Contrary to the canonical
predictions, the regime for which single atoms and solitons co-exist
survives the thermodynamic limit~(\ref{eq:limit}), thus indicating
that the general assumptions of the Landau hypothesis do not directly
apply to the Lieb-Linger model when treated in the MCE. The only way
to on average obtain an $N$-particle  
soliton is for very negative $E_{\rm total} < E_0(N-1)$ and thus $E_{\rm
  total}/E_0( N)\to 1$
in the limit~(\ref{eq:limit}).}
\label{table:overview}
\end{table}
\end{center}

In order to gain a better understanding of what would happen in an
experiment within the co-existence region 
\begin{equation}
0\gtrapprox E_{\rm  total}>E_0(N-1),
\end{equation}
 we will
distribute a negative energy that is proportional to the ground state
energy:
\begin{equation}
\label{eq:gammadef}
E_{\rm total} \equiv \gamma |E_0(N)|, \quad 0> \gamma \ge -1 .
\end{equation}
The number of possibilities to distribute $N$ atoms among up to $N$
solitonlets for the distinct energy eigenstates~(\ref{eq:AllEigenvalues}) is given by the
number partitioning problem which asymptotically reads~\cite{AbramowitzStegun1984}
\begin{equation}
\label{eq:NumberPartitioning}
\Omega_{\rm number~partitioning}(N) \sim\frac{\exp\left(\pi
\sqrt{2/3}\sqrt{N}\right)}{4N\sqrt{3}}.
\end{equation}
When there is no kinetic energy, all energies
of the type given in Eq.~(\ref{eq:AllEigenvalues}) lie between 0 and
the ground-state energy as $(\sum_j N_j)^3\ge \sum_j N_j^3$ for $N_j\ge 0$. Thus, for typical
energies distributing all particles into all possible solitonlets
would give an exponentially growing number of states~(\ref{eq:NumberPartitioning}) --- but this grows
only to the power $\sqrt{N}$ and as we show below is thus not the
leading order contribution.

So far, the above considerations ignore distribution of the kinetic
energy. For $N$ single atoms and large enough kinetic energy $E_{\rm kin}$, the
accessible number of states scales as for the \textit{classical}\/
gas. We note that quantum corrections are negligible as there is no
condensation temperature for non-interacting Bose-gases in a
one-dimensional translationally invariant wave-guide. Thus, all possible quantum
corrections must vanish in the limit~(\ref{eq:limit}) because the
transition temperature remains finite in the canonical
calculations~\cite{HerzogEtAl2014,Weiss2016}. Hence, the leading
contribution to the total number of configurations
reads~\cite{PathriaBeale2011}  
\begin{equation}
\label{eq:scaling}
\Omega_{\rm 1D~gas}(N) \sim \left(\frac{4\pi L^2m} {3\hbar^2}\frac{E_{\rm kin}}{N}\right)^{N/2}.
\end{equation}
If $E$ is (at least) extensive ($\propto N$), this thus grows
considerably faster with $N$ than  the number partitioning
problem~(\ref{eq:NumberPartitioning}).

Outside the crossover region there are no differences between the
canonical and micro-canonical ensemble [see text below Eq.~(\ref{eq:CNsq})], thus we can use the canonical
predictions as a basis for our micro-canonical calculations. This
confirms that the most probable outcome consists of solitonlets of
size $N_r=1$ at high enough total energies. For positive total
energies, all particles can be in such a state. However, for negative
total energies this is no longer the case.

The large exponential growth~(\ref{eq:scaling}) clearly suggests that we should have
as many single atoms as possible. Thus, the negative total energy should be
carried by one large soliton and there should be many solitonlets of unit
size carrying kinetic energy:
\begin{equation*}
E_{\rm total} = E_0(N_1) +\sum_{j=1}^{N-N_1}\varepsilon^{(j)}_{\rm kin}.
\end{equation*}
It would of course be possible to increase the kinetic energy by
adding a further larger solitonlet and reducing the number of single unbound
atoms. But the scaling of Eq.~(\ref{eq:scaling}) and the canonical treatment of the
high-temperature phase~\cite{HerzogEtAl2014,Weiss2016} clearly shows that these configurations are not
statistically relevant.
Thus, the leading contribution to the number of possible
configurations is
\begin{equation}
\Omega(N) \sim \left(\frac{4\pi L^2m} {3\hbar^2}\frac {E_{\rm total}-E_0(N_1)} {N-N_1}\right)^{(N-N_1)/2}
\end{equation}
Taking the leading order approximation $E_{\rm
  total}=E_0(N_1^{(0)})\propto N_1^{(0)} [(N_1^{(0)})^2-1]\sim (N_1^{(0)})^3$,
\begin{equation}
\label{eq:N1}
{N_1^{(0)}} \sim \left(-\gamma\right)^{1/3}N, \quad  0> \gamma \ge -1 ,
\end{equation}
and then using
\begin{equation}
N_1=N_1^{(0)} + \delta N_1,
\end{equation}
we can show that $\delta N_1 /N_1$ is indeed small for negative
$E_{\rm total}$. With $E_{\rm total}-E_0(N_1) = -E_0'(N_1^{(0)})\delta
N_1>0$ the ``maximum finding condition''
\begin{equation*}
\frac{\partial \ln[\Omega(N)]}{\partial \delta N_1 } = 0 
\end{equation*}
then implies in leading order
\begin{equation*}
\frac{N-N_1^{(0)} - \delta N_1}{\delta N_1} - 
\ln\left(\frac{4\pi L^2m} {3\hbar^2}\frac {-E_0'(N_1^{(0)})\delta N_1}
  {N-N_1^{(0)} - \delta N_1}\right)\simeq 0.
\end{equation*}
In the limit~(\ref{eq:limit}) and for $E_{\rm total} \propto E_0(N)$, $E_0'(N_1^{(0)})$ is
independent of $N$ (as is $m$). All other quantities (including $L$) are
proportional to $N$. This yields the leading-order
behaviour~\cite{maple}
\begin{equation}
\delta N_1 =\frac{ \displaystyle \left( 1-\sqrt [3]{-\gamma} \right)N} { {W}
 \left( \,{\frac{ \displaystyle{N}^{4}\pi \,{g_{\rm 1D}}^{2}m^2\left( 
 \left( -\gamma \right) ^{2/3}+\gamma \right) }{\displaystyle 6{\hbar}
^{4} {\varrho_0}^{2} \left(1- \sqrt [3]{-\gamma}\right) }} \right) +1 }
\sim \frac{ \left( 1-\sqrt [3]{-\gamma} \right)N}{2 \ln(N)}
\end{equation}
in the limit~(\ref{eq:limit})  where $\varrho_0 = N/L$ and $Ng_{\rm
  1D}$ are constants. The Lambert $W$ function $W(x)$ solves the equation
  \mbox{$W(x)\exp[W(x)] = x$}.

Repeating the above calculation for $E_{\rm total}>0$ we have as our
starting point
\begin{equation}
\label{eq:0}
N_1=1,\quad \gamma >0,
\end{equation}
 and we have again that $\delta N_1\to 0$ in the limit~(\ref{eq:limit}).

\begin{figure}
\includegraphics[width=\linewidth]{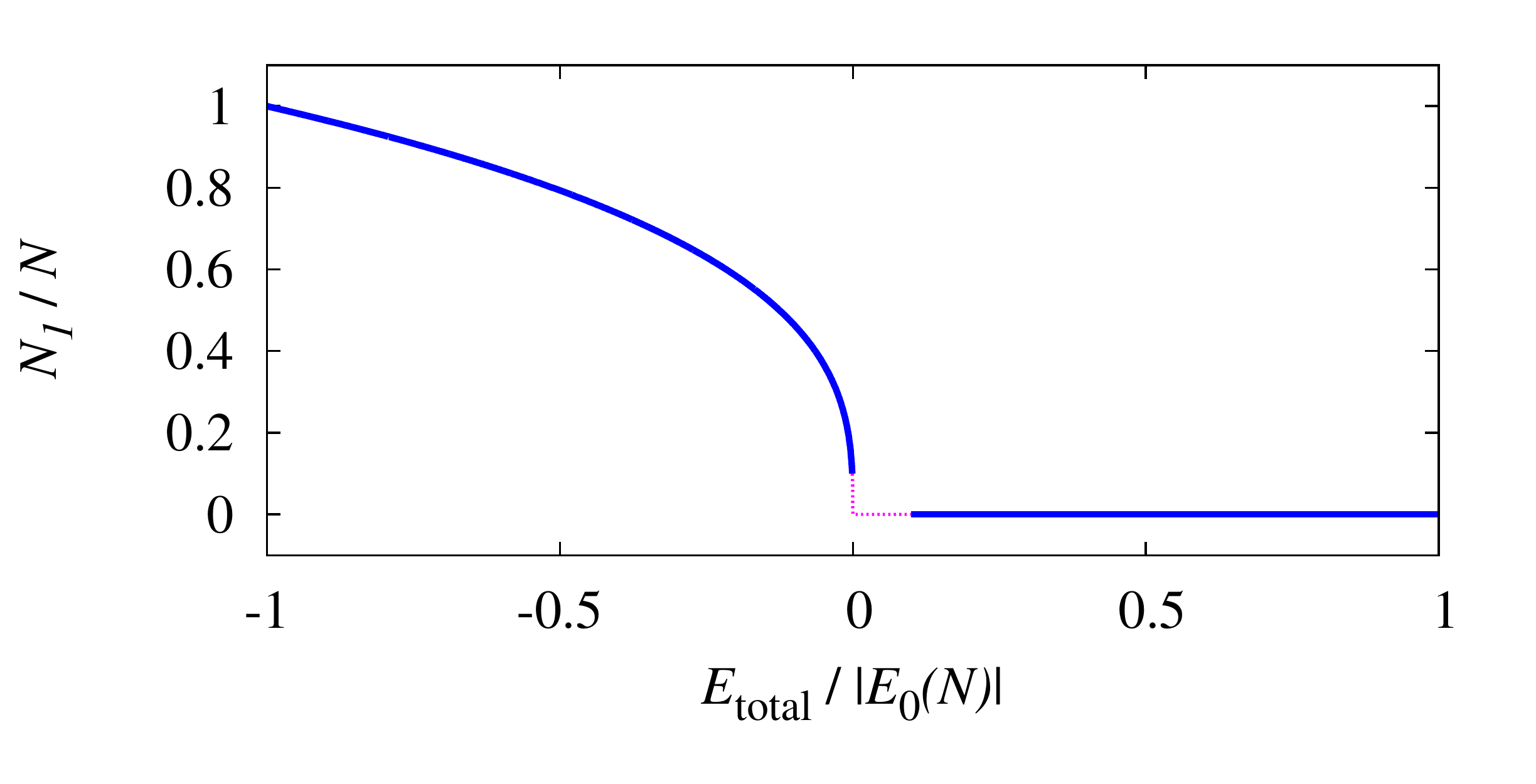}
\caption{Fraction of atoms that are in one big soliton as a function of total
  energy $E_{\rm total}$, Eqs.~(\ref{eq:N1}) and (\ref{eq:0}) as predicted by
  the MCE when approaching the limit~(\ref{eq:limit}). For high energies all atoms are unbound, for
  negative energies all atoms that are not in the largest soliton are
  unbound. A detailed description of the regime $E_{\rm total}\approx
  0$ (thin dashed line) is still an open question. Within the CE, outside a small
  temperature range $\delta T$ that vanishes in the
  limit~(\ref{eq:limit}), only either an $N$-atom soliton or $N$
  unbound atoms can exist in thermodynamic equilibrium~\cite{HerzogEtAl2014,Weiss2016}. }
\label{fig}
\end{figure}
Figure~\ref{fig} shows the size of the largest soliton as
predicted by the MCE. Parameters near $E_{\rm total}\approx
  0$ are consistent with
soliton trains~\cite{HaiEtAl2004,CarrBrand2004,StreltsovEtAl2011}
which were observed in the experiment of Ref.~\cite{StreckerEtAl2002}: for,
say, five solitons of equal size, the total ground state energy would be only
$1/25$ of $E_0$ (in the absence of kinetic energy). 

To conclude, the statistical ensembles MCE and CE are not equivalent for  the Lieb-Liniger model
with attractive interactions. The violation of ensemble equivalence is indicated by anomalous (canonical) energy
fluctuations. We note that the existence of anomalous fluctuations --- that have been discussed
for condensate fluctuations~\cite{SvidzinskyScully2006,GajdaRzazewski1997,Hauge1970}, both for non-interacting Bose-Einstein
condensates~\cite{WeissWilkens1997,GrossmannHolthaus1997a,IdziaszekEtAl1999}
and for interacting
models~\cite{GiorginiEtAl1998,MeierZwerger2001} ---
has been questioned for interacting equilibrium
systems~\cite{Yukalov2005,Yukalov2005b}.

The Landau hypothesis suggests that
a large soliton and single unbound atoms cannot co-exist. While this is
valid in the CE~\cite{HerzogEtAl2014,Weiss2016}, in the MCE
co-existence takes place --- the assumptions on which the Landau
hypothesis is based are valid for many models but
not the Lieb-Liniger model with attractive interactions. This is not
the only curious property of the translationally invariant attractive
Lieb-Liniger model: it is one of the examples for which  a 1D
finite-temperatures phase transition
exists~\cite{HerzogEtAl2014,Weiss2016}; another example can be found
in Ref.~\cite{AleinerEtAl2010}.

As
state-of-the-art experiments arguably are best described by an
isolated system without contact to a heat bath,  neither
anomalous nor normal energy fluctuations occur --- but co-existence of
the two phases does. Using a temperature to
describe bright solitons would require a heat bath and is thus
questionable for state-of-the-art bright soliton experiments in
BECs~\cite{KhaykovichEtAl2002,StreckerEtAl2002,CornishEtAl2006,MarchantEtAl2013,MedleyEtAl2014,McDonaldEtAl2014,NguyenEtAl2014,EverittEtAl2015,MarchantEtAl2016,LepoutreEtAl2016}. Given
that the temperature barely changes over the energy regime relevant
for the existence of bright solitons, it seems to be easier to simply
approximately give it as the transition temperature~\cite{Weiss2016,HerzogEtAl2014} and
calculate it for the experimentally relevant parameters, the result
being practically identical for 73\% 
of all particles occupying one large soliton or, say, 42\%. As the
corresponding energy difference is rather large, getting from one to
the other requires more experimental effort than the tiny temperature
difference might suggest. This behaviour is similar to water freezing
or ice melting without any change of temperature.  For bright solitons in the presence of a harmonic trap,
classical field theory  has been used to investigate the statistics
of an attractive 1D Bose gas~\cite{BieniasEtAl2011}, an approach that might help to
investigate the open question about the statistics near $E_{\rm total}\approx 0$.

The data presented in this paper
 are available online~\cite{WeissGardinerGertjerenken2016Data}.

\acknowledgments

We thank T.~P.~Billam, L.~D.~Carr, Y.~Castin, J.~Tempere and T.~P.~Wiles
for discussions.
C.W.\ and
S.A.G.\ thank the UK Engineering and Physical Sciences Research Council (Grant No.\ EP/L010844/1) for funding. 
B.G.\ thanks the European Union for funding through
FP7-PEOPLE-2013-IRSES Grant Number 605096.

%


\begin{thebibliography}{55}
\bibitem [{foo()}]{footnoteWeissGardinerGertjerenken2016}%
  \BibitemOpen
  \href@noop {} {}\bibinfo {note} {To define sufficiently large $L$, we should
  look at the size of the relative wave function of a two-particle soliton
  [Eq.~(\ref{eq:Psi0})], yielding: $\exp[-m |g_{\rm 1D}| L/(2\hbar^2)]\ll
  1$.}\BibitemShut {Stop}%
\end{thebibliography}

\end{document}